# Quantum in the media: A content analysis of frames in Dutch newspapers

Aletta Lucia Meinsma[1,2]*, Thomas Rothe[1,2]*, W. Gudrun Reijnierse[3], Ionica Smeets[2], Julia Cramer[1,2]

[1] Leiden Institute of Physics, Faculty of Science, Leiden University, Leiden, The Netherlands
[2] Department of Science Communication & Society, Faculty of Science, Leiden University, Leiden, The Netherlands
[3] Department of Language, Literature and Communication, Vrije Universiteit Amsterdam, Amsterdam, The Netherlands

* These two authors contributed equally to this work

**Abstract:** Quantum technology is expected to have an impact on society. Scientists warn against the use of certain frames because they may create barriers to effective science communication. We studied 385 Dutch newspaper articles for the use of these frames. Newspapers commonly explained quantum concepts when mentioning quantum technology. They also regularly framed quantum technology as beneficial and enigmatic, often in prominent positions of the articles. The economic development/competitiveness frame, the social progress frame and the risk frame were less common. Although these frames are only potential barriers, we encourage journalists to weigh them when communicating about quantum technology.

## 1 Introduction

In October 2019, Google claimed to have reached a milestone. They argued they had built a so-called quantum computer that could perform a task in 200 seconds, when – according to them - the world's best supercomputer would take around 10,000 years to complete the task (Arute et al., 2019). News media worldwide paid attention to Google's achievement, including Dutch newspapers.

*Het Parool*, for instance, wrote about possible benefits (Van Unen, 2019)[1]:

> "According to Google, the possibilities are endless in the long term. Think of connecting the data points from which weather forecasts are drawn up at lightning speed, or predicting changes in climate."

*De Telegraaf*, on the other hand, ended their article with concern ('Geheimschrift', 2019):

> "It [i.e. the quantum computer] offers many possibilities, but also potential problems. The encryption, which secures our e-mail traffic, can be cracked in the blink of an eye. And if I were the secret service, I would already start thinking of an alternative to my secret code."

The media coverage of Google's achievement illustrates that quantum technologies, which include quantum computers, are communicated to non-expert audiences in different ways. Quantum technologies are emergent technologies that use quantum physics principles and are categorized into the domains of quantum computing & simulation, quantum communication and quantum sensing & metrology (Stichting Quantumdelta NL, 2020). Most of these technologies are still in their infancy, but it is expected that once they mature, they will start to impact society at large (e.g., European Quantum Flagship, 2020; Stichting Quantumdelta NL, 2020; Vermaas et al., 2019, 2022). Therefore, it is important already at this early stage to consider public engagement with quantum technology, for example to ensure that the technology is built in a socially robust way (Roberson et al., 2021).

The different ways in which quantum technology is communicated to non-expert audiences can impact public engagement in diverging ways. On the one hand, *Het Parool*'s statement that quantum computing can accurately forecast the weather has been called "really far-fetched" (Ezratty, 2022, p. 8). Although it is feared that such hyped up promises will result in a decline in public trust in scientists (Ezratty, 2022), Roberson (2020) suggests that they may also help by raising awareness which may subsequently spark new discussions. On the other hand, the 'quantum computing as a threat' narrative, which *De Telegraaf* mentioned, could also affect public engagement. According to Seskir et al. (2023), this narrative, without presenting a realistic timeline or information on how organizations are already actively working on dealing with the threat, could place time restrictions on potential public engagement and deliberation activities.

As quantum technologies are expected to impact society in the future, there is a role for science communicators and journalists in the process of public engagement with quantum technology. In this paper, we quantitatively examine how quantum science and technology are framed in Dutch newspapers. The theoretical concepts on which our study is based are covered in detail in the next section.

## 2 Theory

Most members of the public become acquainted with scientific information through science-news articles published in (online) media (Schäfer, 2017). In the current online era, despite the emergence of new media platforms such as blogs, social networking sites and video sharing sites, traditional news media continue to play an important role (Weimann & Brosius, 2017). In the Netherlands, for instance, both online and print newspapers are a frequently used source through which citizens interact with information about science and technology (European Commission, Directorate-General for Communication, 2021; Rathenau Instituut, 2021).

As newspapers and other forms of traditional news media emphasise certain news, for instance through the amount of coverage, they can impact what the public considers to be important topics (e.g., Lou et al., 2019). This is known as first level agenda setting (McCombs & Shaw, 1972). In addition to this first level, agenda setting theory also includes a second level (e.g., Scheufele & Tewksbury, 2007; Weaver, 2007; Weimann & Brosius, 2017). While the first level is concerned with *which* topics are discussed in the media, the second level is concerned with *how* the media communicate about those topics (Weaver, 2007). For instance, when discussing a given topic, media outlets can present a positive or negative framing of it by focussing either on the benefits or the risks involved in the issue at hand (e.g., Chuan et al., 2019; Lewenstein et al., 2010; Strekalova, 2015;

Veltri, 2013). This positive or negative framing can also influence people's attitudes towards the issue (e.g., Achterberg, 2014; Cobb, 2005; Druckman & Bolsen, 2011).

The influence of news media is particularly important in the case of emergent technologies (Scherrer, 2023; Scheufele & Lewenstein, 2005), as this is likely the first exposure people have to such a technology. News media coverage for emergent technologies usually seem to follow a bell-shaped curve of object salience: it starts off with a growing amount of coverage followed by a decline (Lewenstein et al., 2010; Nisbet et al., 2003; Veltri, 2013). Previous content analyses examined the news media's coverage of emergent technologies, such as nanotechnology (Lewenstein et al., 2010; Strekalova, 2015), AI (Chuan et al., 2019) and stem cells (Nisbet et al., 2003). The results of these studies show that the news media in general paint a positive picture when reporting on these technologies. The emphasis is, for instance, on economic development, business opportunities and social progress. At the same time, attention is also paid to the risks of the technologies. For example, a content analysis of nanotechnology in the Spanish news media showed that controversies were reported early on (Veltri, 2013). A content analysis of AI in the US news media found that risks were covered less but in more depth than the benefits (Chuan et al., 2019).

Quantum technology is an important emergent technology currently under development. Quantum technology holds the potential to impact society at large once it arrives (e.g., Stichting Quantumdelta NL, 2020; Wehner et al., 2018). As with any technology, this poses both benefits as well as risks for our society. For example, quantum computers have the potential to design new types of materials and molecules that could save or extend lives through drug discovery (Busby et al., 2017; Outeiral et al., 2021), as well as enable new forms of modern warfare that could fall in the hands of terrorist groups (Vermaas et al., 2019). Scientists warn that certain ways of framing quantum technology can create barriers to effective science communication, because it could hinder public engagement (Seskir et al., 2023; Vermaas, 2017) and public trust (Grinbaum, 2017). At the same time, there is also a plea for sufficient attention to reflect on the impact of quantum technology on society (Roberson et al., 2021).

In terms of barriers for effective public communication about quantum, Vermaas (2017), for instance, argues that framing quantum as enigmatic could hinder public understanding of quantum technology and subsequent engagement in societal dialogues to explore the implications of it. Furthermore, according to Seskir et al. (2023), framing quantum technology in terms of having to win a race poses a barrier to participatory efforts with quantum technology between different stakeholder groups. In a military context, for instance, it can lead to research having to be kept secret for certain groups. Thirdly, Grinbaum (2017) states that popular media do not explain underlying quantum physics concepts when mentioning quantum technology, which he argues could influence the public's trust in quantum technology.

In addition to the barriers mentioned above, Roberson et al. (2021) argue that the ways in which quantum technology can impact society for the better should also receive sufficient attention in media coverage. The ways in which quantum technology might improve or solve problems in people's lives (i.e., the social progress frame) should be reflected on. This also means that both the risks as well as the benefits of quantum technology should be reflected on, thereby providing a balanced perspective.

In a recent content analysis of 501 TEDx talks, Meinsma et al. (2023) studied the prevalence of the different quantum-related frames described above. Results of their analysis showed that, while the spooky and enigmatic frame occurred in about a quarter of the talks, the quantum race frame was hardly present. Contrary to what had been suggested in the literature, relatively many TEDx talks contained quantum physics concept explanations. Regarding the balanced perspective, the benefit frame greatly outnumbered the risk frame, while reference to the social progress frame was scarce.

News media likely reach a more diverse group of citizens than TEDx talks, but it is not yet known how news media frame quantum science and technology. In this study we therefore investigate the framing of quantum technology in Dutch newspaper articles. Our research questions are as follows:

> RQ 1: How do journalists frame quantum science and technology in Dutch national newspapers?
> a. How often do journalists frame quantum science and technology as spooky and enigmatic?
> b. How often do journalists frame quantum science and technology in terms of economic development / competitiveness?
> c. How often do journalists explain fundamental quantum concepts when mentioning quantum technology?
> d. How often do journalists frame quantum science and technology in terms of social progress?
> e. How often do journalists frame quantum science and technology in terms of benefits?
> f. How often do journalists frame quantum science in terms of risks?

Moreover, news articles have a specific structure, such that the most important information is shared first, while the remaining text presents less important information (Angler, 2017). This implies that frames positioned in the beginning of news articles (e.g., in the head) are the most prominent. Magusin (2017) highlights three features of heads that make prominent frames worth studying. First of all, heads are often the places that readers tend to read, more often than the full article itself. Secondly, the information in the head guides readers towards the facts presented in the article. And finally, heads rely on cultural knowledge that is believed to be widespread in society, and therefore they may influence the dominant discourse more than the rest of the article. We argue that in addition to the head, the subhead and lead also contain important information (Angler, 2017), and therefore we ask:

> RQ 2: What percentage of the frames that journalists placed in a prominent location are the spooky and enigmatic frame, the economic development / competitiveness frame, the explanation of a fundamental quantum concept when mentioning quantum technology, the social progress frame, the benefit frame, and the risk frame?

## 3 Methods

*3.1 Sample collection*

To answer the research questions, we collected a sample of Dutch newspaper articles with quantum science and technology content. Our data collection method is shown in Figure 1. We used the search string "quantum* OR kwantum*" in the Nexis Uni database (*LexisNexis. Nexis Uni*, n.d.) and set the search window from 01-01-2009 to 31-12-2021. Articles written by the six major Dutch newspapers as listed in the rankings of Nationaal Onderzoek Multimedia (NOM) Dashboard 2022-I were included, namely: *Algemeen Dagblad, De Telegraaf, De Volkskrant, NRC, Trouw,* and *Het Parool*. The search returned a total of 2,553 articles.

In the second step, duplicate articles were deleted via a hybrid automatic-manual process. Details of this process can be found in section A1 in the Appendix and in the referenced source code[2]. It resulted in a total of 2,240 unique articles. Afterwards, we selected the articles with sufficient quantum science and technology content for our study. The reasons for discarding an article are included in Table A1 in the Appendix. The primary reason (n = 599) for excluding an article was due to using the search string in a company or product name, e.g., references to the Dutch company "Kwantum" or the James Bond film "Quantum of Solace". The two first authors of this paper checked the article selection on a 20% random sample of the dataset, which resulted in an acceptable level of agreement ($\alpha = 0.830, 92.4\%$; Krippendorff, 2004). In total, we discarded 1,542 articles which left us with a dataset of 698 articles. The metadata of all these articles were obtained through Nexis Uni (*LexisNexis. Nexis Uni*, n.d.), which included the newspaper brand, the section in which the article was published, the date of publication, the article headline, the name of the author and the word count of the article. Based on formulas for standard error and confidence intervals (Neuendorf, 2017), we drew a random sample of 385 articles for our analysis.

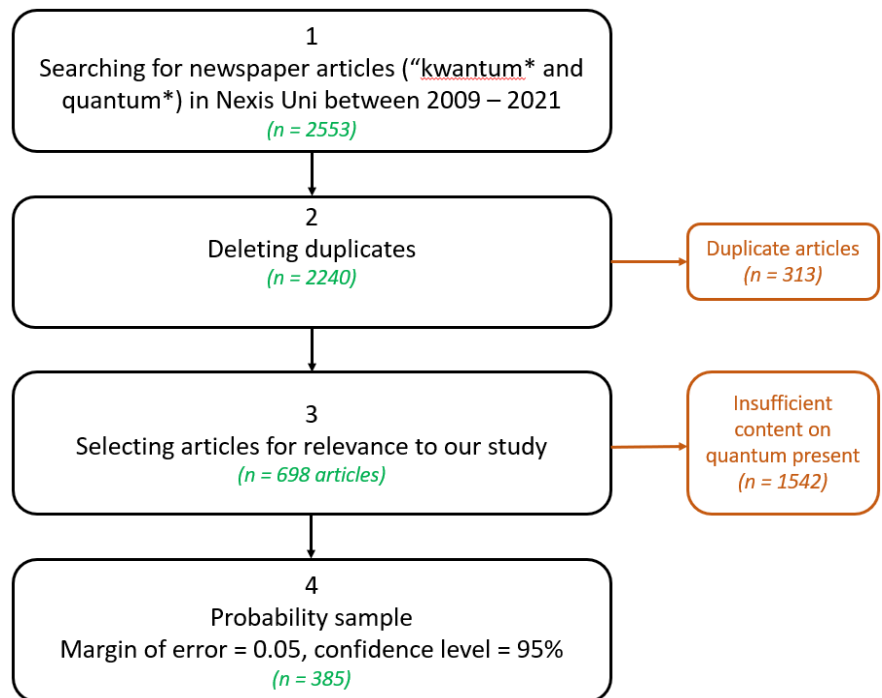

*Figure 1. Data collection method*

*3.2 Codebook*

The complete codebook can be found in Appendix B and was based on the codebook from Meinsma et al. (2023). Descriptives were obtained which include an identification of whether quantum science and/or technology was the main topic of the article and the presence of a mystical viewpoint. A mystical viewpoint denotes ideas that do not adhere to established scientific paradigms and that link quantum physics with spirituality, religion and consciousness, amongst others. Additionally, we identified the article type, where we distinguished between 1) news reports or features, 2) opinion pieces, columns or letters, 3) interviews, and 4) reviews of a product (e.g., of the film 'Ant-Man and the Wasp', in which quantum physics plays a role) or announcements of an upcoming event (e.g., of the Dutch TV show 'DWDD University', where quantum physics would be discussed). We also coded the presence of a quantum technology indicator, which is a term that includes the word 'quantum' and belongs to one of the quantum technology application domains. If applicable, we wrote down the quantum technology indicator. We additionally searched for quantum science applications.

Next, we coded the frames referring to quantum science and/or technology. If a frame was found, the sentence that reveals the frame was copied into the coding sheet. The spooky and enigmatic frame was identified when 'quantum' was associated with 'spooky' or 'enigmatic' or a synonym of those terms. Secondly, the economic development / competitiveness frame was found when a news article made reference to the effect that quantum science and technology can have on the economy, and/or when the article highlights the competitive side on a local, national or global level (see Nisbet, 2009). Thirdly, we identified the presence of an explanation of three types of fundamental quantum concepts: superposition, entanglement and contextuality. We checked for the articles that make reference to quantum technology (i.e., articles with a quantum technology indicator) whether these explain these fundamental quantum concepts. An explanation of superposition includes that a particle can be in multiple states at the same time; of entanglement that two particles share a quantum state, meaning that it does not make sense to discuss those particles as separate entities anymore; and lastly, contextuality is considered a harder concept (see Jaeger, 2019), which we operationalized as that performing a measurement irreversibly affects a quantum state.

Additionally, we coded the frames referring to a balanced perspective on quantum science and/or technology. The social progress frame was present when an article emphasised how quantum science and technology can solve problems or improve people's lives (see Nisbet, 2009). The benefit frame was identified when either a positive evaluation of quantum science and technology was given, quantum science and technology was said to have advantages over something else (e.g., it was attributed to being faster, safer, better, etc.), or when a specific reference was made to how quantum science and technology will benefit a particular field. The risk frame appeared when concerns about quantum science and technology were highlighted. Finally, frames that were placed in the head, subhead or lead were coded as prominent.

*3.3 Coding procedure and reliability*

To determine the reliability of the codebook, the two first authors coded a random selection (n = 54) of the articles within the sample independently from one another. Overall, the codebook was reliably applied by the two coders, and in cases of disagreement, the coders reached a consensus. For an overview of the Krippendorff's alpha values and percentage agreements per code, please see Table A2 in the Appendix. With the discussion in mind, one of the coders proceeded with coding the entire sample.

*3.4 Analysis plan*

The analysis involved calculating the number of times $n$ a code occurred, its sample proportion $p$ and its confidence interval. As we drew a probability sample of 385 articles, these confidence intervals were 95% confidence intervals with a 5% margin of error, meaning we have a 95% confidence that the true population proportion is within 5% of the sample proportion. Before calculating the confidence intervals, we first checked the assumption of at least 15 occurrences and 15 non-occurrences of a code such that $mp \geq 15$ and $m(1-p) \geq 15$, where $m$ is the sample size and $p$ is the sample proportion (*Basic Statistics*, n.d.). If the assumption was not met, we calculated the exact Clopper-Pearson confidence interval instead (indicated with a * in the Results section), which is a more conservative measure (*Epitools - Calculate Confidence Limits for a Sample Prop ...*, n.d.).

# 4 Results

*4.1 Descriptive data*

From the 385 articles that our sample contains, most articles were published in 2014 and 2018 (in both years: n = 41, 10.6%), and the least in 2009 (n = 14, 3.6%). As can be seen in Figure 2, there are spikes in coverage. Figure A1 in the Appendix shows that the curve of the total number of articles in the complete dataset (N = 698 articles) resembles the curve in Figure 2.

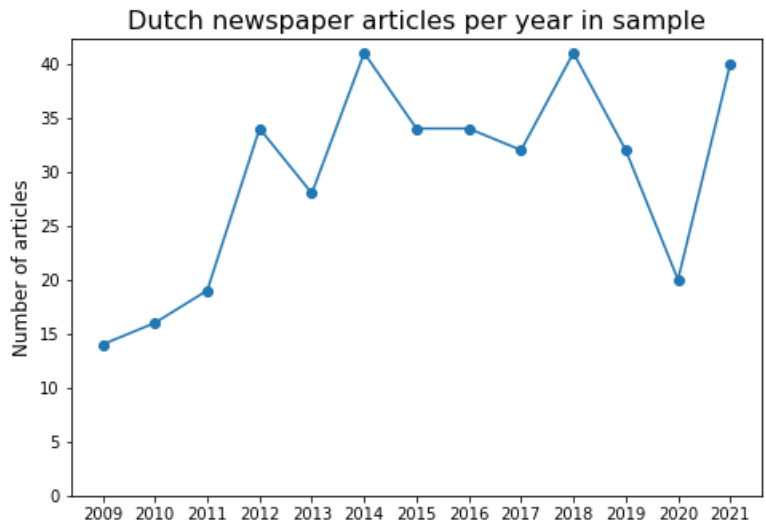

*Figure 2. Frequency of articles by year of publication (N = 385)*

Table 1 gives an overview of the number of articles per newspaper brand in our sample. *NRC* published the majority of articles, followed by *De Volkskrant*. For an overview of the distribution of articles per year per newspaper in our sample, please see Figure A2 in the Appendix. In addition, we identified 118 unique authors in our sample, but in 60 articles the authors name was missing. Similar to Kristensen et al. (2021), the top ten authors contributed largely as they together wrote 51.9% of the articles in our dataset (n = 200).

**Table 1**
*The distribution of articles per newspaper brand (N = 385)*

| Newspaper | Number of articles | Percentage | 95% CI | Number of words |
|---|---|---|---|---|
| *NRC* | 147 | 38.2% | [0.333, 0.430] | 158,967 |
| *De Volkskrant* | 130 | 33.8% | [0.290, 0.385] | 123,973 |
| *Trouw* | 54 | 14.0% | [0.106, 0.175] | 60,565 |
| *Algemeen Dagblad* | 12 | 3.1% | [0.014, 0.049]* | 10,487 |
| *De Telegraaf* | 27 | 7.0% | [0.045, 0.096] | 32,799 |
| *Het Parool* | 15 | 3.9% | [0.020, 0.058] | 7,984 |
| *Total* | *385* | *100%* | | *394,775* |

*Note.* The * indicates the exact Clopper-Pearson confidence interval was calculated.

The majority of articles consisted of news reports or features (n = 234, 60.8%), followed by opinion pieces, columns or letters (n = 60, 15.6%), reviews of a product or announcements of upcoming events (n = 51, 13.2%) and interviews (n = 40, 10.4%). The section that articles most often appeared in was the science section (n = 176, 45.7%) and most articles were published on Saturdays (n = 215, 55.8%).

A little less than half of the articles (n = 170, 44.2%) had quantum science and/or technology as their main focus. Furthermore, we found 178 articles (46.2%) with a quantum technology indicator (i.e., terms that include the term 'quantum' and belong to a quantum technology application domain). Most of these articles mentioned the quantum computing & simulation domain (n = 153, 39.7%), followed by the quantum communication domain (n = 33, 8.6%) and only 2 articles mentioned applications in the quantum sensing & metrology domain (0.5%).

A mystical viewpoint was found in 10 articles (2.9%). An example of such a viewpoint appeared in an interview with a theologist, who pleas for connection and solidarity (Huttinga, 2017):

> "I find it comforting and telling that quantum physics shows us the same thing: everything is completely intertwined and connected - already at the level of the electron. In the universe, everything is mysteriously connected to everything in every possible way. It is up to us to tune into that."

*4.2 Frames*

Table 2 shows the prevalence of the frames that we analysed.

First of all, the spooky and enigmatic frame appeared in 24.2% of the articles in our sample (n = 93). An example of such a frame is "The theory behind quantum mechanics is bizarre and counterintuitive. […] There is nothing weirder than quantum mechanics" (Schenk, 2018). In this example, the terms "bizarre" and "weird" are indicative of the spooky and enigmatic frame as they are semantically similar to the terms "spooky" and "enigmatic".

Secondly, the economic development / competitiveness frame appeared in 8.6% of the articles in our sample (n = 33). An example is "NSA fears European lead in race for 'qubits' " (Hond, 2014), where competitiveness is highlighted through the word "race".

Thirdly, to establish whether articles include an explanation of a quantum physics concept when referring to quantum technology, we analysed the articles with a quantum technology indicator (n = 178) for including an explanation on superposition, entanglement and/or contextuality. Results show that 50.6% of the articles with a quantum technology indicator (n = 90) explained at least one of these three concepts. Out of the three concepts, superposition is explained most often (see Figure A3 in the Appendix). An example of a quantum concept explanation is: "If you apply the elusive properties of quantum physics to classical bits, you take the step to the qubit: an information carrier that can be not only zero or one, but also zero and one at the same time, something that physicists call superposition" (Hal, 2017).

**Table 2**

*Frequency table of the newspaper articles comparing the spooky and enigmatic frame, the economic development/competitiveness frame, the quantum concept explanations for articles with a quantum technology indicator, the benefit frame, the risk frame and the social progress frame.*

| Frame | Total number of articles | Percentage | 95% CI |
| --- | --- | --- | --- |
| Spooky and enigmatic | 93 | 24.2% | [0.199, 0.284] |
| Economic development / competitiveness | 33 | 8.6% | [0.058, 0.114] |
| Quantum concept explanations for articles with quantum technology indicator (n = 178) | 90 | 50.6% | [0.432, 0.579] |
| Social progress | 13 | 3.4% | [0.018, 0.057]* |
| Benefit | 128 | 33.2% | [0.285, 0.380] |
| Risk | 21 | 5.5% | [0.032, 0.077] |

*Note.* Multiple frames can occur in one article. The * indicates the exact Clopper-Pearson confidence interval was calculated.

In only 13 articles (3.4%) reference was made to the fact that quantum would mean something *good* for society and should be developed and deployed in such a way. An example of the social progress frame is: "According to her, quantum technology is going to revolutionise society. It can provide solutions to global issues in climate, energy, healthcare and security" (Van onze correspondent, 2016). The example focuses on quantum technology as a solution to major problems that our society currently faces.

To examine the balanced perspective of quantum technology, we quantified the occurrence of the benefit frame and the risk frame. First of all, the benefit frame appeared in 33.2% of the articles in our sample (n = 128). An example is (Wayenburg, 2014) :

> "The promises are great: with control of quantum information you could build quantum computers that calculate faster than all current computers, because they can analyse many billions of variants of the problem at the same time. Quantum information also offers the possibility to transmit information in an non-eavesdropping manner. And then there are probably even more applications that have not yet been thought of."

By stating that "the promises are great" the author gives a positive evaluation of quantum computers. In addition, by using the word "faster" for comparing quantum computers to current computers, the author emphasises an advantage of quantum computers.

Finally, the risk frame appeared in only 5.5% of the articles in our sample (n = 21). An example of a risk frame is (Brugh, 2016):

"Imagine that everything you email, that you bank online, that you store on your computer, is no longer secure. [...] Peter Schwabe (35), cryptographer at Radboud University Nijmegen, is seriously concerned about that scenario. Because with the imminent arrival of quantum computers, which promise unprecedented computing power, this becomes a real danger."

The phrases "seriously concerned" and "a real danger" in relation to the arrival of quantum computers indicate the risk frame. Furthermore, this example mentions a specific field that is being impacted by the risk: the digital security / privacy field. We did an additional analysis to gain more insight into these specific fields that are mentioned to be benefited or harmed by quantum technology (see figure A4 in the Appendix). Our analysis showed that the digital security and privacy field was most often mentioned to be impacted by quantum science and technology, both in terms of causing benefits as well as causing risks to the field.

In addition to the occurrence of frames in the entire article, we also examined whether occurring frames were placed in prominent locations of the news articles. We only included news reports and features in our analysis, as these are the articles that typically have a structure with the most important information first, followed by less important information. From the articles classified as news reports or features (n = 234), 62.0% (n = 145) includes at least one of the frames that we investigated. In total, we found 61 occurrences of one of the frames in a prominent position. Table 3 gives an overview of the number of times that a frame was placed in a prominent location, and its percentage compared to the total number of prominent frames. The benefit frame was most often placed in a prominent location, followed by the spooky and enigmatic frame.

**Table 3**

*Frequency table of the prominent frames in news reports and features (n = 234). The percentage given is with respect to the total number of prominent frames (n = 61).*

| Frame | Total number of times the frame is prominent | Percentage compared to total number of prominent frames (n = 61) | 95% CI |
|---|---|---|---|
| Spooky and enigmatic | 15 | 24.6% | [0.138, 0.354] |
| Economic development / competitiveness | 6 | 9.8% | [0.037, 0.202]* |
| Quantum concept explanations for articles with a quantum technology indicator (n = 132) | 11 | 18.0% | [0.094, 0.300]* |
| Social progress | 1 | 1.6% | [0.000, 0.088]* |
| Benefit | 25 | 41.0% | [0.286, 0.533] |
| Risk | 3 | 4.9% | [0.010, 0.137] |

*Note.* Multiple prominent frames can occur in one article. In total, 49 articles put at least one of the frames in a prominent position. The * indicates the exact Clopper-Pearson confidence interval was calculated.

Additionally, we compared the number of times a frame was put in a prominent location with the number of times that frame occurred in total in news reports and features. We found that if a frame is present, it is usually placed in a prominent position in a quarter of the news reports and features, as percentages ranged from 22.2% (economic development / competitiveness frame) to 30% (risk frame), with the exception of the quantum concept explanations (15.5%) and the social progress frame (8.3%).

## 5 Discussion

This study examined how quantum physics and technology were described in Dutch newspapers during the period 2009-2021. We quantified the occurrence of six frames relevant to the setting of quantum science and technology.

*5.1 First level agenda setting: how often is quantum technology written about?*

Both in the fully coded sample (385 articles) and in the total dataset of 698 articles with quantum science and technology content we find that the typical bell-shaped curve of salience for emergent technology is not (yet) visible (such as for nanotechnology in the American and Spanish news; Lewenstein et al., 2010; Veltri, 2013). Overall, we see an upward trend interrupted by several dips, including one in 2020 that may be related to Covid-19. As quantum technology is in an early stage of development, it is possible that we are currently at the start of the bell-shaped curve of salience.

Between 2009 and 2021, the six major Dutch newspapers wrote an average of around 30 articles per year with content about quantum science and technology. This number seems relatively small compared to the number of research outcomes in the Dutch media in general (Hijmans et al., 2003) and compared to the prevalence of other physics disciplines (Kristensen et al., 2021). It may thus be that the public has not yet been largely exposed to quantum science and technology and may not know much about it yet. Further empirical research should look into whether people indeed know little about quantum science and technology and do not yet consider the topic important, as dictated by first-level agenda setting.

*5.2 Second level agenda setting: how is quantum science and technology written about?*

We identified the occurrence of six frames related to quantum science and technology. The first frame we identified, the spooky and enigmatic frame, was present in almost a quarter of the articles (24.2%). This is comparable to the frames' 23% occurrence in TEDx talks (Meinsma et al., 2023). Although the frame is not dominant, this potential barrier to effective science communication is thus apparent in Dutch newspaper articles. Because of its regular occurrence, the frame could have implications for people's perceived understanding of quantum science and technology. Further empirical research should investigate whether the spooky and enigmatic frame is prominent in the minds of Dutch newspaper readers, and whether it hinders engagement, as argued by Vermaas (2017). We know from a British study that in a public dialogue exercise around quantum science and technology participants did not mention quantum to be spooky or weird (Busby et al., 2017). If this were also the case for the Dutch situation, this raises questions about the influence of such a frame, given its regular occurrence.

Second, the economic development / competitiveness frame appeared in 8.6% of the articles in our sample of Dutch newspaper articles. Compared to TEDx talks where the frame appeared in 5% of the talks (Meinsma et al., 2023), this percentage is slightly higher, but it is still relatively small. The trend that coverage of an emergent technology focuses on its economic benefits (e.g., Chuan et al., 2019; Lewenstein et al., 2010; Nisbet et al., 2003) is therefore less clear for quantum technology in Dutch newspapers. Given that the focus on competition could be a barrier to public engagement (Seskir et al., 2023), it is encouraging that the economic development / competitiveness frame is not dominant. Its relatively low prominence might cause it to be not very salient in the minds of Dutch newspaper readers, and therefore not very influential for the way people think about quantum science and technology.

Of the articles with a quantum technology indicator (n = 90), 50.6% contained an explanation of at least one of the concepts we studied: superposition, entanglement, and contextuality. Our finding is again similar to the finding in TEDx talks, where 54% of talks with a quantum technology indicator included a statement of superposition, entanglement and/or contextuality (Meinsma et al., 2023). This percentage is higher than expected based on Grinbaum's (2017) concern that popular media is staying away from explaining quantum physics concepts. Grinbaum (2017) argued that a lack of explanation about underlying quantum concepts when popularising quantum technology has a negative impact on public trust. Our analysis, however, reveals that quantum concept explanations are quite common. If Grinbaum (2017) is correct, it may be that in the Dutch context the presence of quantum concept explanations positively influences public trust in quantum technology. To fully explore Grinbaum's (2017) concerns, we recommend further research to study the effect that quantum physics explanations have on public trust.

The least occurring frame in our study was the social progress frame. Only 13 articles (3.4%) mentioned quantum science and technology in the context of solving problems or improving people's lives. This is even less than for TEDx talks (Meinsma et al., 2023), where 7% of the talks included the social progress frame. This is a surprising finding, given that previous research has shown that the social progress frame often appears in discussions about emergent technology (Nisbet, 2009). For instance, Chuan et al. (2019) found that in US newspapers reporting on AI the benefits of AI were mainly emphasized through reference to improving human life or well-being. Interestingly, in a study investigating public attitudes towards quantum technology, Merbel et al. (2023) found that citizens in a neighbourhood of Leiden (The Netherlands) largely agreed with the statement that quantum technology would improve their lives. When asked about the sources people used to gather information about quantum science and technology, Merbel et al.'s (2023) participants primarily mentioned news, museums and media. This is unexpected given our finding that the social progress frame is not very prominent in Dutch newspaper articles, and it raises questions about the relationship between specific attributes in the media (such as the presence or absence of a social progress frame) and people's attitudes towards quantum.

While the social progress frame was hardly present, the benefit frame appeared frequently in our sample of Dutch newspaper articles. A total of 128 articles (33.2%) contained a benefit frame. This is similar to the occurrence of benefits in TEDx talks (34%, Meinsma et al., 2023) and in line with coverage of other emergent technologies, with news media generally painting a positive picture when reporting on these technologies (Chuan et al., 2019; Lewenstein et al., 2010; Nisbet et al., 2003; Strekalova, 2015; Veltri, 2013). Based on our findings, we expect that Dutch newspaper readers generally have a positive image of quantum science and technology. Previous research into the effects of the benefit frame on public perception of new technologies has shown that the frame can have a positive effect on acceptance (e.g., Achterberg, 2014; Cobb, 2005; Druckman & Bolsen, 2011). Follow-up research should show whether the benefit frame is indeed salient in the minds of Dutch newspaper readers, and whether this in turn gives them a positive perception of quantum science and technology.

Compared to the salience of the benefit frame, the risk frame is much less common. Only 21 articles (5.5%) mentioned concerns about quantum science and technology, which is again similar to the finding in TEDx talks (5%, Meinsma et al., 2023). The fact that the frame rarely occurs is consistent with previous research into other emergent technology (Chuan et al., 2019; Lewenstein et al., 2010; Nisbet et al., 2003; Strekalova, 2015). As a wider sense of how quantum technology constitutes harm is lacking, the wider public good narrative as advocated by Roberson et al. (2021) does not really seem to exist. Based on our findings, we expect that the risks of quantum technology are not very prominent in Dutch newspaper readers' minds.

Finally, we identified the occurrence of frames in news reports and features that were positioned in a prominent location. The frame that was most often put in a prominent location (the head, subhead or lead) was the benefit frame (41.0%) followed by the spooky and enigmatic frame (24.6%). As readers tend to read these frames most often because of their location in the article, these may be more influential than the other frames. A limitation of our study is that we did not take the other types of articles into account, where the most important information may be in a different position. Further research could investigate this.

*5.3 Overall comparison TEDx talks and Dutch newspaper articles*

The occurrence of the six frames that we identified in Dutch newspaper articles is overall very similar to their occurrence in TEDx talks (Meinsma et al., 2023). This was unexpected, as the format (presentations vs news articles), language (English vs Dutch), content creator (non-expert or expert speaker vs journalist), and audience (local communities and web users who watch the talk on YouTube vs Dutch newspaper readers) differ. Given that the TEDx talks transcripts contained around 2,400 words and the news articles in our sample contained an average of 956 words, it would be interesting to investigate whether the format has an effect on how the frames are presented. For example, explanations of quantum concept explanations might be explained in more depth in TEDx talks, due to the larger word count. Further research could delve deeper into this.

An interesting difference between the two formats is the appearance of a mystical viewpoint. Though 73 TEDx talks (15%) made reference to such a viewpoint, only 10 news stories (2.9%) connected quantum physics to concepts like spirituality, religion, and consciousness. Perhaps, most journalists want to ensure that the ideas they present about quantum science and technology fit inside widely accepted scientific paradigms.

*5.4 Limitations*

Our study used a predefined set of frames to code the data. With such a top-down approach, interesting other frames may be overlooked. By using a bottom-up approach, these missed frames are revealed. We advise further research to make use of an inductive coding procedure to examine which frames emerge.

Another limitation of our study is the fact that only a small number of authors have made a major contribution to the framing of quantum science and technology in Dutch newspaper articles. There are likely to be self-reinforcing effects, with journalists looking at each other's articles, work they have previously written or other communications about quantum science and technology.

Finally, although the concept is more nuanced, we operationalized contextuality as the idea that a measurement irreversibly affects a quantum state. We have limited our codings to only address this aspect of a measurement because of the concept's complexity. Further qualitative research could examine which quantum concepts are all described and how deeply they are discussed in popular communication.

*5.5 Practical implications*

We would like to advise journalists and science communicators to consider how to present quantum science and technology to a general audience. As news articles may potentially influence people's perceptions of, and subsequent attitudes towards, quantum technology, there is an important role for journalists in the process of public engagement with quantum technology. We encourage journalists and science communicators to already consider the barriers to effective science communication that scientists warn about, even though these barriers are only potential barriers. In this way, the communication about quantum science and technology might jump from a state of superposition—where it is effective and ineffective at the same time— to a state of effectiveness.

*Notes*
*1. The quotes have been translated from Dutch.*
*2. Code repo available at https://github.com/t-rothe/quantum-in-Dutch-newspapers.*

# Acknowledgements

We acknowledge funding from the Dutch Research Council (NWO) through a Spinoza Grant awarded to R Hanson (Project Number SPI 63-264) and thank Ronald Hanson for this opportunity. This work was supported by the Dutch National Growth Fund (NGF), as part of the Quantum Delta NL programme.

# Appendix A

*A1. The hybrid automatic-manual process to detect duplicates*

To detect duplicates, we used a hybrid automatic-manual process. We defined two articles as duplicates if they were: 1) perfect 1-to-1 copies, 2) a basic version vs. an extended/edited version, 3) a preview vs. the main article, or 4) copies of articles with small changes in individual words or clauses (i.e., not a perfect 1-to-1 copy, but articles with equal content and matching sentences for most words).

For articles to be duplicates, the overlap had thus to be at least one whole paragraph (single matching sentences were not sufficient). Exceptions to this were if two articles shared 1-to-1 paragraphs but both had at least one exclusive paragraph, we kept both in. Also, if two articles were duplicates, but they were published on dates 3 or more months apart, we kept both because to readers these articles could appear to be independent of each other and consequently result in a double salience of the framing effect. Also, the context and relevance of a similar article published on different dates may change over time, so that a later republished article may be perceived differently than on its original publication date. If agency reports are stretched/extended by a newspaper editor, they were also not marked as duplicates since they might contain unique content for the newspaper brand. Finally, articles that were highly similar in length and content and for which most sentences had been paraphrased were both kept in, as paraphrasing could have affected the framing content.

To make sure that we detected all types of duplicates, we wrote a script that automatically evaluates the similarity of articles based on both edit- & overlap-distances to cover all the various duplicate types (we used the following similarity metrics: Damerau-Levenshtein Distance, Ratcliff/Obershelp "Distance", Overlap "Distance"). The articles that our script identified as very similar were checked manually. In order to prevent missing duplicates, we chose this minimum value to be on the low side relative to the typical similarity scores that we found for duplicates in 2009 and 2021. As a consequence, the second author still checked hundreds of article pairs manually, but this number was much lower than if we would have performed a full manual duplicate check. Articles were removed from our dataset if they met our definition of a duplicate article. Perfect 1-to-1 copies and articles with small changes in individual words or clauses merely occurred 1) for articles with related brands (e.g. NRC.NEXT and NRC, which we merged into the single code NRC); and 2) for articles from the same news brand but published on different dates (<3 months apart). In the first case, the article from the main brand (e.g. NRC) was kept in the dataset, and for the second case, the article with the latest publication date was kept in. For basic versions vs. extended/edited edition and for a preview vs. the main article, the article with the most words was kept in the dataset.

Further technical details about the automated part of this process can be found in the scripts and accompanying instructions of our code repo: https://github.com/t-rothe/quantum-in-Dutch-newspapers.

*A2. Reasons for discarding newspaper articles as irrelevant or unsuitable for the analysis*

**Table A1**
*Total number of articles discarded from the 2,240 unique articles*

| # | Indicator | Example quote | Total excluded |
|---|---|---|---|
| 1 | Keyword "Quantum" contained in (common) terms unrelated to quantum science / technology, e.g.:<br>- "quantum leap"<br>- "quantum grey"<br>- A 'quantum' as in a quantity of something | "The latest campaign cleverly responds to previous **quantum leaps**: 'The strippenkaart became the OV-chipkaart.'"<br>"He took it for granted that I was talking about the garbage bag gray colour instead of **quantum gray**." | 141 |
| 2 | Keyword "Quantum" used in a proper noun that is unrelated to (quantum) science / technology, e.g. a company name or product name | "**Quantum** of Solace" [movie]<br>"**Quantum** Leben"<br>"He added that the Dutch occupants were in a **Quantum**-minivan belonging to tour company Eco Coaches." | 599 |
| 3 | Metaphorically referring to quantum (concepts) to make a point / explain something else, i.e. without further notice or explanation of (quantum) science / technology. | "**Quantum mechanics** states that light is a wave and a particle at the same time. […] And now I'm actually proposing something similar where people are concerned. Can you experience another person as a fellow human being and as a stranger at the same time?"<br>"In the same week in which the minister says he would like to be a **quantum particle** - minister and scientist in one - …" | 35 |
| 4 | Mentioned a word/proper noun containing the keyword "Quantum" as part of a text/document that only forms a:<br>- TV guide<br>- Table of contents for (news) articles<br>- (Event) Announcements (also for lectures / talks without reasonable discussion of content)<br>- Short independent corrections on earlier articles (e.g. misspellings) | "SUN 5 APR Search for the lowest temperature and visible **quantum phenomena**. Lecture by physicist Dirk Bouwmeester." | 74 |
| | - other short listings of independent and incoherent sentences / words<br>instead of a normal newspaper article. | | |
| 5 | Mentioning / Listing quantum science (concepts), technology as a scientific or technological example without further mention, explanation or discussion of the topic. | "... for example an LED, or another light source like a **quantum well**."<br>"A broad palette: with the A for Alzheimer's, the L for LGBTQIA+, and the Q for **quantum computer**." | 223 |
| 6 | Mentioning scientific instruments / experiments or names of other things with the keyword "Quantum" without mentioning or discussing its relation to quantum science (concepts) or technology | "The rocket carried the so-called X-Ray **Quantum** Calorimeter ..."<br>"Rotational Field **Quantum** Magnetic Resonance" | 54 |
| 7 | Used the topic or a concept of quantum physics / technology as an example to make a point or explain something else, i.e. without further notice of (quantum) science / technology.<br>[Note: 'use' = more extensive / focused than 'mention' or 'listing' ] | "If you also want to describe what happens inside molecules, you have to do **quantum mechanical** calculations. But then you get nowhere - then you can only describe a hydrogen atom." | 241 |
| 8 | Mentioning or listing a person or institution that is related to / works on / knows about quantum science or technology without further mention or discussion of the topic itself. | *"*The science battle between **quantum scientist** Julia Cramer and cognitive neuroscientist Barbara Braams: it will be spectacular.*"*<br>"She studied chemistry in the 1920s at a time when **quantum mechanics** was just emerging." | 109 |
| 9 | Mentioning the topic of quantum science (concepts) or technology to indicate that something unrelated is (not) difficult / (not) complex (to understand).<br>OR<br>Indicating that the author/someone else does not understand quantum science / technology. | "Now Kleine Goos knows as much about the [Sacred] Scripture as about **quantum mechanics**, ..."<br>"I won't bother children with **quantum mechanics** either, because they obviously don't understand anything about it."<br>"How I need a drink, alcoholic of course, after the heavy lectures | 33 |
| 10 | Very shortly mentioning quantum physics in relation to paranormal, consciousness, reality without any explanation of the quantum physics part | "I wonder if there is something of his inner world left in his skull, an energetic **quantum-like** something, in a matter that I cannot observe." | 33 |
| *Total* | | | 1,542 |

*Note.* The quotes have been translated from Dutch.

*A3. Intercoder reliability results*

**Table A2**
*Intercoder reliability results for the different codes that are categorised under descriptives, frames and explanations. The codes with a low agreement (α < .667) are marked in red.*

| | | Amount of times '1' is coded | Krippendorff's α | Percentage agreement |
|---|---|---|---|---|
| **Descriptives** | Article type | - | 0.79 | 88.9% |
| | Main focus | - | 0.69 | 75.6% |
| | Quantum technology indicator | 24 | 0.93 | 96.3% |
| | Quantum computing & simulation | 21 | 0.87 | 96.0% |
| | Quantum communication | 4 | 0.63 | 92.0% |
| | Quantum sensing & metrology | 0 | - | 100% |
| | Mystical viewpoint | 1 | 1.0 | 100% |
| **Frames** | Spooky and enigmatic | 11 | 0.82 | 94.4% |
| | Econ dev / comp | 4 | 0.88 | 98.1% |
| | Social progress | 4 | 0.65 | 96.3% |
| | Benefit | 17 | 0.82 | 92.6% |
| | Risk | 3 | 1.0 | 100% |
| **Explanation** | Superposition | 11 | 0.77 | 92.6% |
| | Entanglement | 4 | 1.0 | 100% |
| | Contextuality | 3 | 0.65 | 96.3% |
| **Prominent** | Spooky and enigmatic | 4 | 0.27 | 75.0% |
| | Econ dev / comp | 2 | 0.53 | 75.0% |
| | Social progress | 1 | - | 100% |
| | Benefit | 6 | 0.84 | 92.9% |
| | Risk | 2 | 1.0 | 100% |

*Note.* These are the intercoder reliability results based on the coding of 20% of a sample that contained 267 articles (n = 54 articles). In the end, because time allowed it, a total of 385 articles were coded. The results section also contains an analysis of the quantum science explanation prominence code. This analysis was done based on a written discussion between the first and second coder.

*A4. Additional figures*

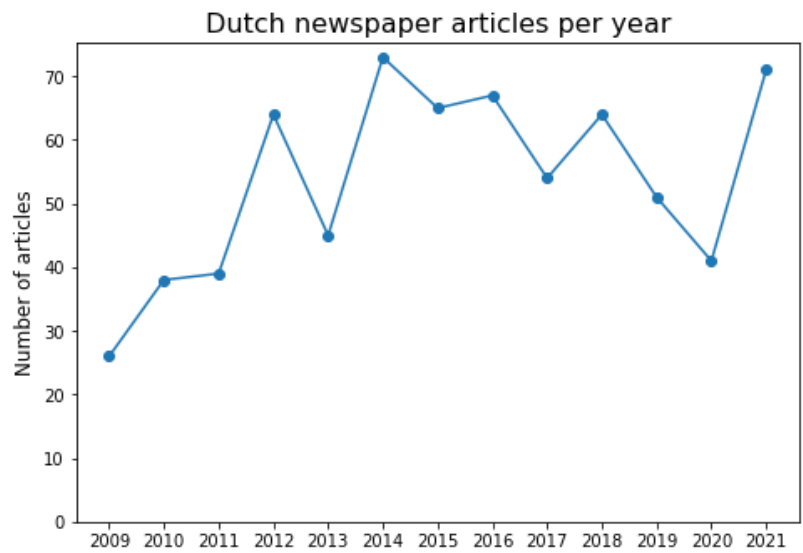

*Figure A1. Number of articles by year of publication in total dataset (N = 698).*

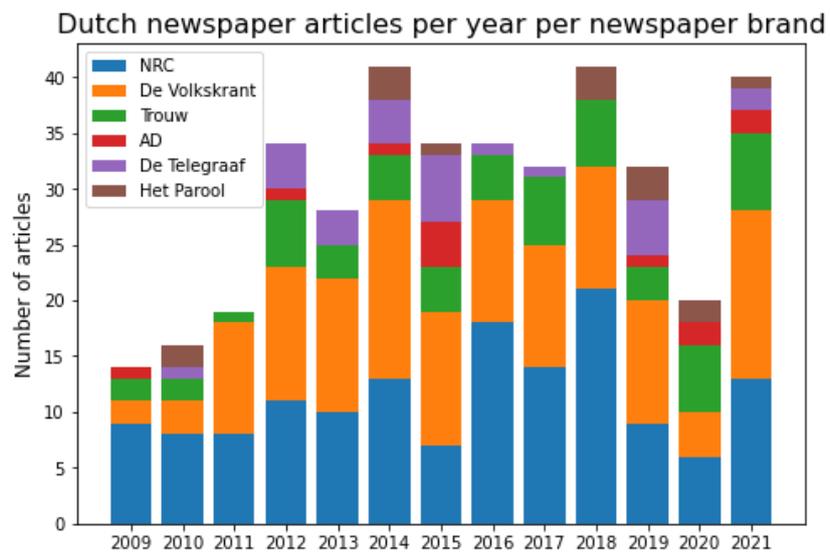

*Figure A2. Number of articles by year of publication per newspaper brand (N = 385).*

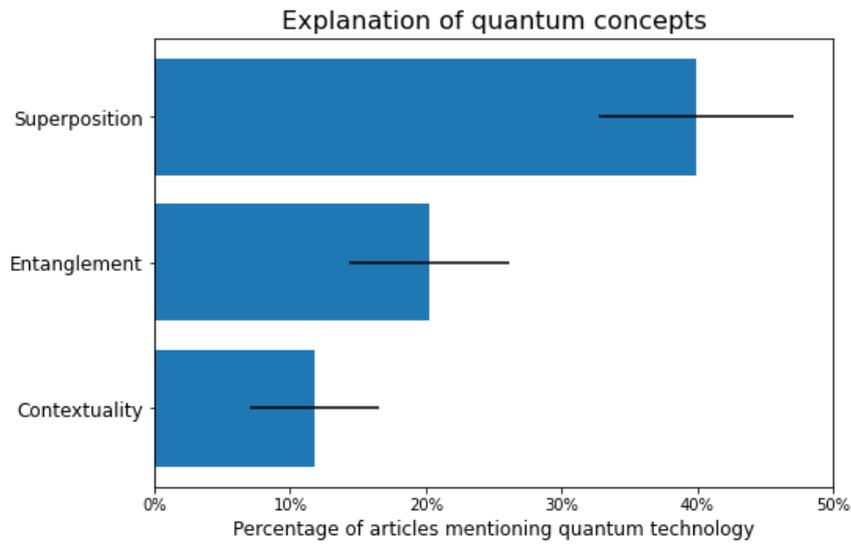

*Figure A3. The percentage of articles that explain superposition, entanglement or contextuality when referring to quantum technology. The error bars are based on the sampling.*

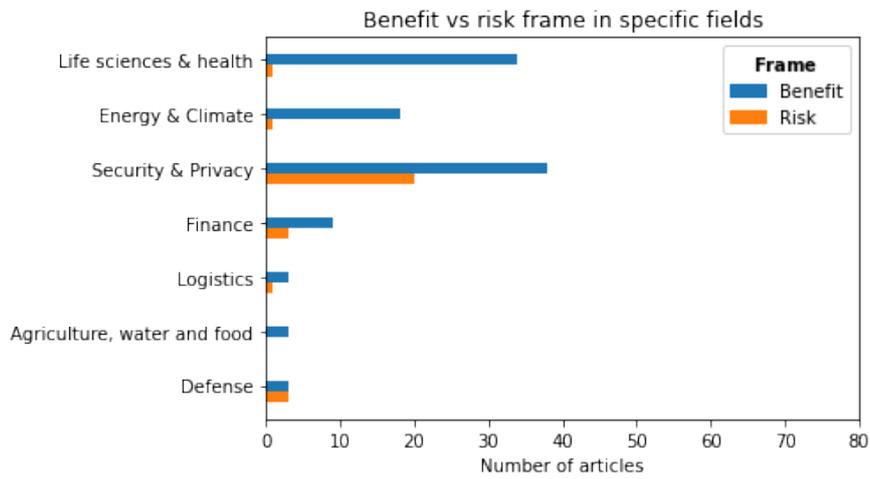

*Figure A4. Number of times a specific field was mentioned in terms of benefits and risks.*

# Appendix B

# Codebook

## Part I: Metadata

| # | Question | Options / Instructions |
|---|---|---|
| 0. | **Who is coding?** | (a) Coder A <br> (b) Coder B |
| 1. | What is the **URN content-ID** of the article **in NexisUni**? | - If still existent, remove the constant ending *"-00000-00"* <br><br> *e.g. XXXX-XXXX-XXXX-XXXX* |
| 2. | What is the **Newspaper Name / Brand**? | (a) NRC; NRC.nl; NRC Handelsblad; NRC.NEXT <br> (b) de Volkskrant; de Volkskrant.nl <br> (c) Trouw; Trouw.nl <br> (d) AD/Algemeen Dagblad; AD/Algemeen Dagblad.nl; <br> (e) De Telegraaf <br> (f) Het Parool |
| 3. | What is the **Publication Year**? | *e.g. 2021* |
| 4. | What is the **Publication Month**? | Jan. => 1, Feb. => 2, Mar. => 3 <br> April => 4, May => 5, June => 6 <br> July => 7, Aug. => 8, Sept. => 9 <br> Oct. => 10, Nov. => 11, Dec. => 12 |
| 5. | What is the **Publication Day** of month? | *e.g. 23* |
| 6. | What is the **Publication Weekday**? | Mon. => 1, Tue => 2, Wed. => 3 <br> Thu. => 4, Fri. => 5, Sat. => 6 <br> Sun. => 7 |
| 7. | What is/are the **Author Name(s)**? | - *Code a "~" if author missing / unclear* <br> - *If multiple authors, list them separate by semicolons ";"* <br><br> *e.g. Dorine Schenk* |
| 8. | In which **Newspaper section** was the article published? | Only use the section metadata provided by NexisUni. <br> Code a "~" if unknown / non-existent in metadata |
|  |  | *e.g. Wetenschap* |
| 9. | Quote the **Article Headline** | *e.g. "Fysici 'teleporteren' informatie via rudimentair quantuminternet"* |
| 10. | What is the **Word Count** of the article body**?** | Use word count from NexisUni metadata *e.g. 583* |

## Part II: Content

|  |  |  |
|---|---|---|
| 11. | What is the **main topic** of the article**?** | - Quantum Nanotechnology: Applications of quantum science (e.g. nanotechnology, QT 1.0, QT 2.0)<br>- Examples of quantum science related topics: quantum cosmology (incl. Big Bang theory and the black hole information paradox), particle physics, high energy physics, photonics, condensed matter physics, nuclear physics, post-quantum cryptography<br>- The main topic is the overarching theme / main trigger of an article.<br>- If an article consists of multiple fully independent sections then focus on the section(s) containing a quantum keyword.<br><br>**(x) Other / Not clear**<br>**(a) Quantum Nanotechnology**<br>**(b) Quantum Science or a related scientific topic** |
| 12. | Which **article type** does the article represent? | - Only deduce the article type from the newspaper section when the type stays ambiguous based on the writing style and other type-unique features.<br><br>**(a) News report / feature**<br>**(b) Opinion / Column / Letter**<br>**(c) Interview**<br>**(d) Review of a product or Announcement of an upcoming event** |
| 13. | Is there a **Holistic Viewpoint** present in the article**?** | **(0) No**<br>**(1) Yes** |
| 14. | Does the article **contain any QT 2.0 indicator?** | "Quantum technology 2.0 indicators [necessarily] include the term 'quantum' and belong to one of the following application domains: quantum computing & simulation, quantum communication, and/or quantum sensing & metrology. " [TEDx codebook] |
| | | For example: quantum sensor; quantum bit; quantum algorithm<br><br>**(0) No**<br>**(1) Yes** |
| **14-Q.** Quote the QT 2.0 indicator in the article | | - If 14. was coded "(0) No", code here a "~".<br>- If multiple indicators present, quote only the first indicator from the headline, highlight, leading paragraph or article body (descending priority)<br>-<br>*e.g. "quantum technology"* |
| **15.** Is **any** kind of **Quantum Technology 1.0 / QT 2.0 / other Application of Quantum Science mentioned?** | | - Only consider 'Yes' if the connection to quantum physics (concepts) is made evident in the article<br>- Also objects or phenomena that only feature/demonstrate quantum-related behaviour are interpreted as quantum applications.<br><br>**(0) No**<br>**(1) Yes** |
| **16.** Which Quantum Technology / other Application of Quantum Science is mentioned? | | - If 15. was coded "(0) No", code here all applications with a -2<br>- Code option I with "1" when the category of an application is ambiguous or when there is doubt about the status as a "quantum application"<br><br>Code each quantum application (A - I) with a 0 (absent) or a 1 (present):<br><br>A => Laser<br>B => MRI scanner<br>C => Smartphone<br>D => Computer<br>E => Nuclear energy<br>F => Quantum computer or quantum simulator<br>G => Quantum network, quantum internet, quantum cryptography<br>H => Quantum sensor<br>I => Other / unsure |
| **16-Q.** Quote the mentioned Quantum application in the "Other / unsure" category. | | - If 15. or option I (Other / unsure) in 16. was coded with "0", code here a tilde "~".<br><br>*e.g. "QLEDs"* |

## Part III: Frames

<table>
<tr><td rowspan="5"><b>Frame 1: Spooky / Enigmatic</b></td><td colspan="2"><i>Frame definition:</i> The article includes "A synonym of spooky or enigmatic that refers to quantum science, a quantum science principle, or a quantum science application." [TEDx Codebook]<br><br>Examples of Dutch synonyms* for spooky [spookachtig]: akelig, geheimzinnig, onheilspellend, ijzingwekkend, griezelig, omineus, ijselijk, creepy, eng, bovennatuurlijk<br><br>Examples of Dutch synonyms* for enigmatic [raadselachtig]: delfisch, enigmatisch, cryptisch, schimmig, geheimzinnig, mysterieus, mystiek, vreemd, omineus, obscuur<br><br>Also include terms which are no clear synonyms but that have a very similar meaning, e.g. 'absurd', 'wonderlijk', 'ongrijpbaar','krankzinnig'.<br><br>(*): https://vandale.nl , https://synoniemen.net en https://www.mijnwoordenboek.nl/synoniem.php</td></tr>
<tr><td><b>17.</b> Can the "Spooky / enigmatic"-frame be identified in the article?</td><td>Only consider 'Yes' (1) whenever the synonym for spooky or enigmatic link to a quantum concept or application of quantum science explicitly.<br><br>(0) No<br>(1) Yes</td></tr>
<tr><td><b>17-Q.</b> Quote the first word or (sub)phrase that reveals the frame.</td><td>- Code a "~" if the frame is absent (17. = 0 = No)<br>- If the frame occurs multiple times, quote only the first occurrence that is found in the headline, highlight, leading paragraph or article body (descending priority)<br><br>e.g. " […] die mysterieuze quantumeigenschappen […]"</td></tr>
<tr><td rowspan="2"><b>Frame 2: Economic development / competitiveness</b></td><td colspan="2"><i>Frame definition:</i> The article transports the idea that "Various parties are in competition to develop quantum nanotechnology, there is a quantum race going on. Quantum nanotechnology will provide economic growth, and will therefore have an impact on all kinds of industries." [TEDx Codebook]. The frame exists if at least one of the two components are found.</td></tr>
<tr><td><b>18.</b> Can the "Economic development / competitiveness"-frame be identified in the article?</td><td>Only consider 'Yes' (1) whenever economic development / competitiveness originates from quantum technologies and/or applications<br>(0) No<br>(1) Yes</td></tr>
</table>

| | | |
|---|---|---|
| | **18-Q.** Quote the first sentence that reveals the frame. | - Code a "~" if the frame is absent (18. = 0 = No)<br>- If multiple sentences apply, quote only the first sentence that is found in the headline, highlight, leading paragraph or article body (descending priority)<br><br>*e.g. "Dankzij Kouwenhovens onderzoek leek Microsoft de concurrentie in een klap voorbij te kunnen streven."* |
| **Frame 3: Social progress** | *Frame definition:* The article transports the idea that "Quantum (w.r.t quantum science and its applications:) would mean something *good* for society and should be developed and deployed in such a way.". The positive tone must therefore be explicit! | |
| | **19.** Can the "Social Progress"-frame be identified in the article? | Only consider 'Yes' (1) whenever social progress originates from quantum technologies or applications<br><br>**(0) No**<br>**(1) Yes** |
| | **19-Q.** Quote the sentence that reveals the frame. | - Code a "~" if the frame is absent (19.= 0 = No)<br>- If multiple sentences apply, quote only the first sentence.<br><br>*e.g. " […] de laatste ontwikkelingen in de quantumtechnologie [...]. Iets waarvan Broer zelf constateert dat dit "[...] in potentie de wereld kan veranderen" ".* |

| | 20. What **kind of causality** is primarily insinuated between QTs and the emergence of social progress**?** | - Code a tilde "~" if the frame is absent (19. = 0 = No)<br>- Code an 'x' (Unclear) if the frame occurs multiple times with inconsistent causality types<br><br>**(x) None / Unclear**<br>**(a) Probabilistic causality**<br>*e.g. "Maar we weten wel dat er een kans is dat ze voor enorme ontwrichting kunnen zorgen. "*<br>**(b) Monocausality**<br>*e.g. "So quantum physics has been responsible for developing technology that define our current day society."*<br>**(c) Sufficient-component causality**<br>*e.g. "Quantum technologie, biotechnolgie en kunstmatige intellligentie zullen er voor zorgen dat iedereen in toekomst toegang krijgt tot goedkope medicatie en gezondsheidszorg."* |
|---|---|---|
| **Frame 4: Benefits & Risks** | We define the "mention of a benefit" of quantum science (applications) in a broader sense as an:<br>- expression of value or (explicit) positive evaluation of quantum science (applications)<br>- comparison of quantum science (applications) with something else for which the quantum science (application) is ascribed with an advantage (e.g. "faster", "safer", …)<br>- concrete mention of an emergent case / situation / characteristic of quantum science (applications) that provides profit / gain towards a specific field. | |
| | 21. Can **any benefits** of quantum technology (or other applications of quantum science) be identified in the article**?** | **(0) No**<br>**(1) Yes** |
| | 21-Q. Quote the sentence that indicates / mentions the benefit | - Code a "~" if there is no benefit mentioned at all (21. = 0 = No)<br>- If multiple benefits mentioned, quote only the first sentence.<br><br>*e.g. "De quantumcomputer kan daarmee in theorie vele malen sneller berekeningen maken [...] voor bijvoorbeeld weer- of klimaatmodellen "* |
| | 22. What **kind of causality** is primarily insinuated between QTs and the emergence of benefit**?** | - Code a "~" if there is no benefit mentioned at all (21. = 0 = No)<br>- Code an 'x' (Unclear) if the frame occurs multiple times with inconsistent causality types |
| | | **(x) None / Unclear**<br>**(a) Probabilistic causality**<br>*e.g. "Zo blijken quantumradars veelbelovend voor het opsporen van 'stealth' vliegtuigen, die voor gewone radar niet goed zichtbaar zijn."*<br>**(b) Monocausality**<br>*e.g. "Qutech en Google zetten daarvoor hun quantumcomputers in, de voorlopers van een rekenbeest dat bepaalde sommen veel sneller kan oplossen."*<br>**(c) Sufficient-component causality**<br>*e.g. "Door deze eigenschap kan een quantumcomputer, mits die voldoende qubits heeft, berekeningen uitvoeren waar gewone computers miljoenen jaren over zouden doen."* |
| | **23.** Which specific **field(s)** are mentioned as being **impacted by the benefit?** | - Code all fields -2 if there was no benefit mentioned at all (21. = 0 = No)<br><br>Code each field (A - G) with a 0 (absent) or a 1 (present):<br><br>A => Medical & Life sciences & health<br>B => Finance<br>C => Logistics<br>D => (Digital) Security & privacy<br>E => Defense<br>F => Energy & Climate<br>G => Agriculture, food and water |
| | **24.** Can **any risks** of quantum technology (or other applications of quantum science) be identified in the article**?** | **(0) No**<br>**(1) Yes** |
| | **24-Q.** Quote the first sentence that indicates / mentions the risk | - Code a "~" if there is no risk mentioned at all (24. = 0 = No)<br>- If multiple risks/disadvantages mentioned, quote only the first sentence of the one that is found in the headline, highlight, leading paragraph or article body (descending priority)<br><br>*e.g. "Een nieuw probleem leveren quantumcomputers ook op."* |
| | 25. What **kind of causality** is primarily insinuated between QTs and the emergence of risk**?** | - Code a "~" if there is no risk mentioned at all (24. = 0 = No)<br>- Code an 'x' (Unclear) if the frame occurs multiple times with inconsistent causality types<br><br>**(x) None / Unclear**<br>**(a) Probabilistic causality**<br>*e.g. "Maar we weten wel dat er een kans is dat ze voor enorme ontwrichting kunnen zorgen. "*<br>**(b) Monocausality**<br>*e.g. "Als bijvoorbeeld de kwantumcomputer er komt, dan betekent dat dat er geen veilige verbindingen meer zijn en onderzeeërs op grote diepte zichtbaar worden."*<br>**(c) Sufficient-component causality**<br>*e.g. "De kwantumcomputer verschijnt aan de horizon en als die techniek doorbreekt, zijn alle online veiligheidsmaatregelen in één klap waardeloos."* |
| | 26. Which specific **field(s)** are mentioned as being **impacted by the risk?** | - Code all fields -2 if there was no risk mentioned at all (24. = 0 = No)<br><br>Code each field (A - G) with a 0 (absent) or a 1 (present):<br><br>A => Medical & Life sciences & health<br>B => Finance<br>C => Logistics<br>D => (Digital) Security & privacy<br>E => Defense<br>F => Energy & Climate<br>G => Agriculture, food and water |
| | 27. Is the risk weakened, marginalized, or contradicted in any way? | - 'Yes' for a single risk is sufficient<br>- The mention of a specific way of mitigation is sufficient<br>- If no risk mentioned at all (24. = 0), then code -2<br>-<br>**(0) No**<br>**(1) Yes** |
| General Framing | A frame is prominent if it is (primarily) noticeable from the headline, highlight or leading paragraph (descending in priority) | |
| | 28. What are the **prominent frames** of the article**?** | - If a specific frame was not present (17., 18., 19., 21. or 24. coded 0), code here for that frame a -2. |

| | | Code each frame (A - E) with a 0 (not dominant) or a 1 (dominant):<br><br>A => Spooky / Enigmatic frame<br>B => Economic Development / Competitiveness frame<br>C => Social Progress frame<br>D => Benefit frame (exclude economic (B) and social (C) benefits)<br>E => Risk frame<br>F => Quantum science explanations*<br><br>*This variable was added later and was not part of the original codebook. |
|---|---|---|

## Part IV: Explanations

| | |
|---|---|
| **29.** Which **quantum concepts** are **mentioned?** | Code each concept (A - C) with a 0 (absent) or a 1 (present):<br><br>A => Superposition[1]<br>B => Entanglement[2]<br>C => Contextuality[3] |
| **30.** Which **quantum concepts** are **explained?** | - If explanations of multiple concepts are entangled within a single statement/sentence, code all the concepts that covered.<br><br>Code each concept (A - C) with a 0 (not explained) or a 1 (explained):<br><br>A => Superposition[1]<br>B => Entanglement[2]<br>C => Contextuality[3] |
| **30-Q-A.** Quote the <u>first</u> sentence that contains an explanation of <u>superposition.</u> | Code a "~" if there is <u>no</u> explanation of superposition (A = 0 in 29.)<br><br>*e.g. "Allereerst 'superpositie', dat ervoor zorgt dat de quantumversie van een bit, een qubit, niet alleen '0' of '1' is, maar ook '0' en '1' gelijktijdig kan zijn. "* |
| **30-Q-B.** Quote the <u>first</u> sentence that contains an explanation of <u>entanglement.</u> | Code a "~" if there is <u>no</u> explanation of entanglement (B = 0 in 29.)<br><br>*e.g. " [...] als twee elementaire deeltjes eenmaal in dezelfde quantumtoestand zijn gebracht: die twee blijven eeuwig* |
| | *verbonden, ook al belanden ze in verschillende uithoeken van het universum."* |
| **30-Q-C.** Quote the first sentence that contains an explanation of contextuality. | Code a "~" if there is no explanation of contextuality (C = 0 in 29.) <br><br> e.g. *" [...] dat is het merkwaardige van quantumtoestanden: ze liggen in het duister tot je ze meet, en als je ze meet is hun informatie verloren."* |

[1] "A particle in a superposition state can be in multiple quantum states at the same time. For example, when an electron is in a superposition state, it can exist in spin states up and down at the same time." (Meinsma et al., 2023)

[2] "Two entangled particles share an extremely strong connection with each other - measuring one of the particles instantly affects the state of the other, even when the particles are separated by a large distance. In other words: entangled particles can only be described by the quantum state for the entire system, and not by their individual quantum states" (Meinsma et al., 2023)

[3] "Contextuality means that "outcomes of measurements [depend] on other measurements on the same system". This means that when performing a measurement on a quantum state, that measurement affects the quantum state irreversibly" (Meinsma et al., 2023). The statement of a measurement being destructive suffices. A sole reference to measurements and/or their probabilistic nature is *not* enough.

Please note that this is the complete codebook applied by the coders. Not all coded variables were reported on in this article. For the sake of transparency, they are listed here.